\documentstyle[preprint,aps,eqsecnum,epsfig]{revtex}
\tightenlines
\def\pmb#1{\setbox0=\hbox{#1}%
     \kern-.025em\copy0\kern-\wd0
      \kern.05em\copy0\kern-\wd0
       \kern-.025em\raise.0433em\box0}

\def\beq{\begin{equation}}
\def\eeq{\end{equation}}
\def\bea{\begin{eqnarray}}
\def\eea{\end{eqnarray}}
\begin{document}
\title{Kinetic-theory approach to low-energy collective modes in nuclei}
\author{ V. Abrosimov$^{\rm a}$, A.Dellafiore $^{\rm b}$ and 
F.Matera$^{\rm b}$}
\address{  $^{\rm a}$ \it Institute for Nuclear Research, 252028 Kiev, 
Ukraine\\
 $^{\rm b}$ \it Istituto Nazionale di Fisica Nucleare and Dipartimento
di Fisica,\\
Universita' di Firenze, Largo E.Fermi 2, ~50125 Firenze, Italy}
%\date{}
%
\maketitle

\begin{abstract}
	Two different solutions of the linearized Vlasov equation for
finite systems, characterized by fixed and moving-surface boundary conditions,
are discussed in a unified perspective. A condition determining the eigenfrequencies	
of collective nuclear oscillations, that can be obtained from the
moving-surface solution, is studied for isoscalar vibrations of lowest
multipolarity. Analytic expressions for the friction and mass parameters
related to  the low-enegy surface excitations are derived and
their value is  compared to values given by other models. Both similarities
and differences are found with respect to the other approaches, however
the close agreement obtained in many cases with one of the
other models suggests that, in spite of some important differences, the
two approaches are substantially equivalent. The formalism based on the Vlasov
equation is more transparent since it leads to analytical expressions that can
be a basis for further improvement of the model.
\end{abstract}
\vspace{.5 cm}
\vspace{.5 cm}

\newpage
\section{Introduction}

	A collisionless kinetic equation with time-dependent mean field
has been used long ago, first by Vlasov and then by Landau, to describe
plasma oscillations in the high-frequency regime \cite{lan}.
More recently Bertsch has proposed to use the same equation to study small-amplitude
oscillations in nuclei \cite{ber}.
 The approach
of Vlasov and Landau assumes translation invariance of the system under study,
extensions of their approach to finite systems have to face the nontrivial
problem of which boundary conditions to impose on the fluctuation of the
sigle-particle density distribution $\delta n({\bf r},{\bf p},t)$.

A solution of the linearized Vlasov equation suitable for describing
giant resonances of various multipolarities in heavy spherical nuclei has been
derived in \cite{bri}. The boundary conditions employed in \cite {bri}
are inspired by the Steinwedel-Jensen (SJ) model of the giant dipole resonance (GDR)
(compressible fluid within fixed boundary, see for example Ref.\cite{rin},
p.558).
 In Ref.\cite{abr} an
approach similar to that of Ref.\cite{bri} has been followed, but with
boundary conditions that are inspired by the Goldhaber-Teller (GT),
rather than by the SJ model. In the GT model of the GDR the 
neutron and proton fluids are assumed to oscillate against each other
without being compressed (see for example Ref. \cite{rin}, p. 558).

In order to adopt such a description, the authors of \cite{abr} had to account
for the reflection of nucleons on the moving nuclear suface (assumed to be
sharp), and this was achieved by modifying the boundary conditions employed
in \cite{bri}.

	In this paper we first recall the approaches of Refs. \cite{bri} and
\cite {abr}, giving a unified discussion of the different solutions
derived in those papers. Then we consider the condition determining the
eigenfrequencies of collective modes that has been derived in Ref. \cite{abr}.
Since in that paper this condition has been studied in detail only for
monopole ($L=0$) oscillations, we turn our attention to the lowest isoscalar
modes with $L>0$. We introduce a low-frequency approximation that allows us
to obtain analytical expressions for the dynamic coefficients (friction
and mass parameters) entering the equation of motion for the surface
oscillations of a system of fermions. These coefficients and the associated
eigenfrequencies of collective oscillations are then compared to the analogous
expressions given by the liquid-drop model (LDM) of Ref. \cite{boh} and by a
more closely related model discussed in Ref. \cite{koo}

\section{One-dimensional example}

	In Ref. \cite{bri} it has been shown that for a spherical system the
solution of the linearized Vlasov equation can be reduced to the solution of
a one-dimensional problem in the effective  potential $U_0 (r) +{{\lambda^2}
\over{2mr^2}}$.

 In this section we briefly compare the approaches of
\cite{bri} and \cite{abr} by assuming a one-dimensional model, our aim is
to make the connection between the two approaches more transparent.
	Thus we consider a one-dimensional system described by an equilibrium
Hamiltonian
\beq
h_0 (x)=\epsilon ={{p^2}\over{2m}}+U_0 (x).
\eeq
The (possibly self-consistent) equilibrium mean field $U_0 (x)$ is assumed
to have the shape of a potential well, so that the motion of particles
is bounded within two classical turning points $x_{1(2)}$ defined by
$U_0(x_{1(2)})=\epsilon$. In the Hartree approximation the relation between
the self-consistent field and the interaction between particles
$V(\mid x-x'\mid)$ is
\beq
\label{mf}
U_0 (x)=\int dx'dp'V(\mid x-x'\mid)n_0 (x',p'),
\eeq
with $n_0 (x,p)$ the single-particle phase-space distribution at equilibrium.  
We assume that at time $t=0$ this system is acted
upon by a weak external field of the type $\delta(t) Q(x)$ which induces a
time-dependent fluctuation of the equilibrium single-particle
distribution: $n_0 (x,p) \to n_0 (x,p)+\delta n(x,p,t)$.
As a consequence the equilibrium mean field also acquires a time-dependence:
$U_0 (x) \to U_0 (x) +\delta U(x,t)$.
It is convenient to take the Fourier transform with respect to time and to
change variables from $(x,p)$ to $(x,\epsilon)$, so that
\beq
\delta n(x,\epsilon,\omega)=\delta n_+ (x,\epsilon,\omega)+
\delta n_- (x,\epsilon,\omega),
\eeq
with $\delta n_{\pm}  (x,\epsilon,\omega)=
\delta n(x,p=\pm\sqrt{2m[\epsilon-U_0 (x)]},~\omega)$.
Then, according to \cite{bri}, the linearized Vlasov equation implies
\bea
\label{sis}
{{\partial \delta n_{+}(x,\epsilon,\omega) }\over{\partial x}}-
A(x,\epsilon,\omega)\delta n_+ (x,\epsilon,\omega)=B_+ (x,\epsilon,\omega)
\nonumber \\
{{\partial \delta n_-  (x,\epsilon,\omega)}\over{\partial x}}+
A(x,\epsilon,\omega)\delta n_-  (x,\epsilon,\omega)=B_- (x,\epsilon,\omega)
\eea
with 
\beq
\label{af}
A(x,\epsilon,\omega)={{i\omega}\over{\sqrt{{2\over m}[\epsilon-U_0 (x)]}}}
\eeq
and
\beq
\label{bterm}
B_{\pm}(x,\epsilon,\omega)= ({{\partial n_0}\over{\partial \epsilon}})[
{{dQ(x)}\over{dx}}+
{{\partial \delta U(x,\omega)}\over{\partial x}}].
\eeq

	The solution of the system (\ref{sis}) is not completely trivial
because the fluctuation of the mean field
\beq
\delta U(x,\omega)=\int dx'dp'V(\mid x-x'\mid)[\delta n_+ (x',\epsilon ',\omega)
+\delta n_- (x',\epsilon ',\omega)]
\eeq
couples the two equations. In the approach of \cite{bri} the solution is obtained
in two steps, first the mean field fluctuation in (\ref{bterm}) is neglected,
thus obtaining a zero-order solution $\delta n^0_{\pm}$
that can be derived explicitly, then the solution with the full inhomogeneous
term (\ref{bterm}) is expressed in implicit form through an integral equation
similar to that given by the random-phase approximation (RPA) for the
particle-hole Green function.

	Of course the zero-order solution $\delta n^0_{\pm}$ depends on the
boundary conditions. The conditions employed in \cite{bri} were
\beq
\label{fs}
\delta n_{+}(x_{1(2)},\epsilon,\omega)=
\delta n_{-}(x_{1(2)},\epsilon,\omega).
\eeq
In this case the solution of the system (\ref{sis}) (neglecting mean-field
fluctuations) becomes
\beq
\label{no}
\delta n^{0}_{\pm} (x,\epsilon,\omega)=e^{\pm i\omega\tau(x)}
[\int_{x_1}^{x}dx' B^0 (x',\epsilon)
e^{\mp i\omega\tau(x')}~~+~~C(\omega)],
\eeq
with $B^0 (x,\epsilon)=
 ({{\partial n_0}\over{\partial \epsilon}})[{{dQ}\over{dx}}]$,
\beq
\tau(x)=\int_{x_1}^x dy {{1}\over{\sqrt{{2\over m}[\epsilon-U_0 (y)]}}}~,
\eeq
\beq
C(\omega)={{e^{i\omega T}\int_{x_1}^{x_2} dx' B^0 (x',\epsilon)
e^{-i\omega \tau(x')}-
\int_{x_1}^{x_2} dx' B^0 (x',\epsilon)e^{i\omega \tau(x')}}\over{1-e^{i\omega T}}}~,
\eeq
\beq
T=2\tau(x_2).
\eeq
(We have simply re-written in more compact form the solution given in
Sect.4 of \cite{bri}. Note that there is a misprint in Eq.(4.13b) of
\cite{bri}, that should read:
$C_2 ={{C_r -C_l exp(-i\omega T/2)}\over{1-exp(i\omega T)}}$ ).

	In Ref. \cite{abr} it has been argued that the effects of moving
boundary can be taken into account, at least in a linearized approximation,
simply by changing the boundary conditions (\ref{fs}). In particular, since
we are interested in the radial motion of nucleons in the effective
potential $U_0 (r) +{{\lambda^2}\over{mr^2}}$, it is sufficient to change the
boundary condition at one turning point (say $x_2$). Thus, instead of the
boundary conditions (\ref{fs}), in \cite{abr} the following boundary
conditions have been used
\bea
\label{ms}
\nonumber
\delta n_{+}(x_{1},\epsilon,\omega)=
\delta n_{-}(x_{1},\epsilon,\omega)~~~~~~~~~~~~~~~~~~~\\
\delta n_{+}(x_{2},\epsilon,\omega)=
\delta n_{-}(x_{2},\epsilon,\omega)+
({{\partial n_0}\over{\partial \epsilon}})F(\omega),
\eea
where $F(\omega)$ is  the Fourier transform of an as yet undetermined
function of time $F(t)$.

	The solution analogous to (\ref{no}) with these new boundary conditions
is
\beq
\label{dno}
\delta \tilde{n}^0 =\delta n^0 +\delta n_{S}^0
\eeq
with
\beq
\delta n_{S}^0(x,\epsilon,\omega) =({{\partial n_0}\over{\partial \epsilon}})
F(\omega)~{{\cos[\omega\tau(x)]}\over{i\sin(\omega T/2)}}.
\eeq

	 We can obtain $\delta n^0 (t)$ by taking
the inverse Fourier transform of $\delta n^0 (\omega)$
\beq
\delta n^0 (t)={1\over{2\pi}}\int_{-\infty}^{+\infty} d\omega e^{-i\omega t}\delta n^0
(\omega),
\eeq
which can be determined by contour integration in the complex $\omega$-plane.
Since we are interested in $\delta n^0 (t)$ for $t>0$, the integration
contour must be closed in the lower part of the complex plane. The first term
on the $r.h.s.$ of Eq. (\ref{no}) is analytic in the whole plane, and the only
contributions to the integral come from the poles of $C(\omega)$ that are at
\beq
\omega= n{{2\pi}\over{T}}-i\eta \qquad (n~{\rm integer},~\eta\to 0^+ ),
\eeq                                                                   
where $1-e^{i\omega T}=0$. Equivalently we can thik of $\omega$ as a complex
variable having a small positive imaginary part $i\eta$.
	The moving-boundary solution is more complicated since the pole
structure of $\delta n^0_S (\omega)$ is not as simple as that of
$\delta n^0 (\omega)$.
Apart from the usual poles at $\omega= n{{2\pi}\over{T}}$, the quantity
$\delta n^0_S (\omega)$ can have additional poles due to $F(\omega)$,
and these extra poles may contribute to the contour integral. The problem of
determining the explicit form of $F(t)$, and consequently the pole structure
of $F(\omega)$, will be discussed in the next section.

\section{Three-dimensional (spherical) case}

	In this section we recall the approach followed in Ref. \cite{abr}
in order to determine the pole structure of the additional term in
the boundary condition (\ref{ms}).

	The approach of \cite{abr} is inspired by the LDM
\cite{boh}, the nuclear mean field is approximated by a square-well type
potential, but, contrary to Ref. \cite{bri}, in this model the external field
is allowed to change also the equilibrium shape of the nucleus.
The
change of shape is parametrized as 
$R(\theta,\varphi,t)=R+\delta R(\theta,\varphi,t)$, with
\beq
\delta R(\theta,\varphi,t)=\sum_{LM} \delta R_{LM}(t) Y_{LM} (\theta,\varphi).
\eeq
In the LDM a surface deformation of this kind generates a
fluctuation of the pressure at $r=R$ that is given by (cf. Eq.(6A-57) of
\cite{boh})
\beq
\label{dp}
\delta P({\bf r},t)\mid_{r=R}=\sum_{LM}C_L {{\delta R_{LM}(t)}\over{R^4}}
Y_{LM} (\theta,\varphi).
\eeq
The restoring force parameter $C_L$ used in \cite{abr} contains only
the contribution of the surface energy,  so
\beq
\label{cl}
C_L \approx (C_L )_{surf} =\sigma R^2 (L-1)(L+2),
\eeq
and $\sigma$ is the surface tension parameter that can be obtained from
the mass formula:
\beq
\sigma\approx 1 {\rm ~MeV~fm}^{-2}.
\eeq

	In Ref.\cite{abr} the link with kinetic theory was established by putting
\beq
\label{dpig}
\delta P({\bf r},t)\mid_{r=R}=\int d{\bf p} p_r v_r [\delta n({\bf r},{\bf p},t)
 -({{\partial n_0}\over{\partial \epsilon}})\delta U ({\bf r},t)]~\mid_{r=R}.
\eeq
This is a kind of self-consistency condition and its physical meaning is that
the nuclear surface is required to behave as a free surface.
The integral on the $r.h.s.$ is the radial-radial element of 
the momentum flux tensor $\delta \Pi_{rr}({\bf r},\omega)$ for the nuclear
quantum liquid \cite{lif}.
The radial velocity $v_r (r)$ is
\beq
v_r (r)=\sqrt{{2\over m}[\epsilon-U_0 (r)-{{\lambda^2}\over{2mr^2}}]},
\eeq
while the radial momentum is $p_r =mv_r$.

	Combining Eqs.(\ref{dp}) and (\ref{dpig}), and taking the
Fourier transform with respect to time gives
\beq
\label{fun}
\sigma R^{-2}\sum_{LM} (L-1)(L+2)\delta R_{LM}(\omega)Y_{LM} (\hat{\bf r})=
\delta \Pi_{rr}({\bf r},\omega)\mid_{r=R} .
\eeq
This equation can be used to determine the poles of the
function $F (\omega)$ appearing in the three-dimensional generalization
of Eq.(\ref{ms}), that is, the eigenfrequencies of collective nuclear
vibrations.
Multiplying both sides of Eq.(\ref{fun}) by
$Y_{LM}^* (\hat{\bf r})$ and integrating over solid angle gives
\beq
\label{sigma}
\sigma R^{-2}(L-1)(L+2) \delta R_{LM}=\delta\Pi^{LM}_{rr}(r,\omega)
\mid_{r=R}.
\eeq
The $r.h.s.$ is the multipole component of the integral in Eq.(\ref{dpig}), it
is given by a generalization of Eq.(B.6) of \cite{bri}:
\bea
\label{b6}
\delta\Pi^{LM}_{rr}(r,\omega) & = & {{(4\pi)^2}\over{(2L+1)}}~{1\over 2r^2}
\int d\epsilon \int d\lambda \lambda p_r \\
\nonumber
& \{ & \sum_{N=-L}^{L}[\delta n_+ (LMN,\epsilon\, \lambda\, r,\omega)
+\delta n_- (LMN,\epsilon\, \lambda \, r,\omega)]Y_{LN}^*
({\pi\over 2},{\pi\over 2}) \\
\nonumber
& - & ({{\partial n_0}\over{\partial \epsilon}}){{2L+1}\over{4\pi}}
\delta U_{LM}(r,\omega)\}.
\eea

The quantities $\delta n_{\pm}(LMN,\epsilon\, \lambda\, r,\omega)$
satisfy the system of differential equations (\ref{sis}) with \cite{bri}
\beq
A(N,\epsilon\, \lambda\, r,\omega)
={{i\omega}\over{v_r (r)}}-i{{N}\over{v_r (r)}}{{\lambda}\over{mr^2}}
\eeq
replacing Eq.(\ref{af})and (there are some misprints in Eq.(5.12 of \cite{bri})
\beq
\label{bf}
B_{\pm}(LMN,\epsilon\, \lambda\, r,\omega)=
({{\partial n_0}\over{\partial \epsilon}})
[({{\partial}\over{\partial r}}\pm i{{N}\over{v_r (r)}}{{\lambda}\over{mr^2}})
(Q_{LM}(r)+\delta U_{LM}(r,\omega))]Y_{LN}({\pi\over 2},{\pi\over 2})
\eeq
instead of Eq.(\ref{bterm}).
	Like in the one-dimensional example, the solution of the linearized
Vlasov equation (\ref{sis}) can be studied in two steps. At first we can
neglect, both in Eqs.(\ref{b6}) and (\ref{bf}), the mean-field fluctuation
$\delta U_{LM}$ that couples the two equations, thus obtaining a zero-order
approximation that may be a starting point for deriving the more complete
solution that takes into account $\delta U_{LM}$. In this paper we investigate
only the zero-order approximation and show that even at this level some
interesting results about low-energy isoscalar collective modes can be
obtained.

	The zero-order solution analogous to Eq.(\ref{no}) (fixed surface) is
\bea
\label{nolmn}
\nonumber
\delta n^{0}_{\pm} (LMN,\epsilon\, \lambda\, r,\omega) & = &
e^{\pm i[\omega\tau(r)-N\gamma (r)]}
[\int_{r_1}^{r}dr' B_{\pm}^0 (LMN,\epsilon\,\lambda\, r')
e^{\mp [i\omega\tau(r')-N\gamma(r')]} \\
& + & C(LMN,\epsilon\, \lambda,\omega)],
\eea
with
\beq
B_{\pm}^0 (LMN,\epsilon\, \lambda \, r)= ({{\partial n_0}\over{\partial \epsilon}})
[{{dQ_{LM}}\over{dr}}\pm i {{N}\over{v_r (r)}}{{\lambda}\over{mr^2}}
Q_{LM}]Y_{LN}({\pi\over 2}{\pi\over 2}),
\eeq
\beq
\tau(r)=\int_{r_1}^r dy {{1}\over{\sqrt{{2\over m}[\epsilon-U_0 (y)-
{{\lambda^2}\over{2my^2}}]}}}~,
\eeq
\beq
\gamma(r)=\int_{r_1}^r dy {{1}\over{\sqrt{{2\over m}[\epsilon-U_0 (y)-
{{\lambda^2}\over{2my^2}}]}}} {{\lambda}\over{my^2}}~,
\eeq
\bea
\nonumber
C(LMN,\epsilon\, \lambda,\omega) & = &
\{e^{i[\omega T-N\Gamma]/2}\int_{r_1}^{r_2} dr
B_+^0 (LMN,\epsilon\, \lambda \,r)
e^{-i[\omega \tau(r)-N\gamma(r)]} \\
& - & \int_{r_1}^{r_2} dr B_{-}^0 (LMN,\epsilon\, \lambda \,r)
e^{i[\omega \tau(r)-N\gamma(r)]}\} \{1-e^{i[\omega T-N\Gamma]}\}^{-1} ~,
\eea
\beq
T=2\tau(r_2 ) \qquad\Gamma=2\gamma(r_2 )~.
\eeq

	The solution analogous to Eq.(\ref{dno}) (moving surface) instead, reads
\bea
\label{nolmnti}
\nonumber
\delta \tilde{n}^{0}_{\pm} (LMN,\epsilon\, \lambda\, r,\omega) & = &
\delta n^{0}_{\pm} (LMN,\epsilon\, \lambda\, r,\omega)\\
& + & ({{\partial n_0}\over{\partial \epsilon}})
e^{\pm i[\omega\tau(r)-N\gamma (r)]}
F(LMN,\epsilon\, \lambda,\omega)~~{1\over{2i}}{1\over{\sin({{\omega T-N\Gamma}\over{2}}})}.
\eea

	One more step is necessary in order to get the final result of
Ref. \cite{abr}: the functions $F(LMN,\epsilon\lambda,\omega)$
must be related to the radial velocity of the surface (cf. Eq. (3.9) of
\cite{abr}) through
\beq
F(LMN,\epsilon\,\lambda,\omega)=
i\omega 2p_r (\epsilon\, \lambda,R) \delta R_{LM}(\omega)
Y_{LN}({\pi\over 2}{\pi\over 2}).
\eeq

Replacing this expression into Eq.(\ref{nolmnti}) and recalling that
$\delta n=\delta n_+ +\delta n_-$, gives
$\delta \tilde{n}^0 =\delta n^0 +\delta n_{S}^0$ ,
with
\beq
\delta n_{S}^0 =2({{\partial n_0}\over{\partial \epsilon}})
\cos[\omega\tau(r)-N\gamma(r)]~
\omega p_r (\epsilon\, \lambda,R) \delta R_{LM}(\omega)
Y_{LN}({\pi\over 2}{\pi\over 2}){1\over{\sin({{\omega T-N\Gamma}\over{2}}})}.
\eeq
Combining this equation with Eqs.(\ref{sigma}) and(\ref{b6}) and neglecting
the mean-field fluctuation $\delta U_{LM}$ in (\ref{b6}) gives the following
expression for $\delta R_{LM}(\omega)$ in zero-order approximation:
\bea
\label{dr0}
& \delta & R_{LM}^0 (\omega)  = \\
\nonumber
& &{{{{(4\pi)^2}\over{(2L+1)}}~{1\over 2R^2}
\int d\epsilon \int d\lambda \lambda p_r 
\sum_{N}\delta n^0 (LMN,\epsilon\, \lambda\, R,\omega)
Y_{LN}^* ({\pi\over 2},{\pi\over 2})}\over{\sigma{{(L-1)(L+2)}\over{R^2}}
-{\omega\over R^2}{{(4\pi)^2}\over{(2L+1)}}~
\int d\epsilon ({{\partial n_0}\over{\partial \epsilon}})
\int d\lambda \lambda p_{r}^2 
\sum_{N}\mid Y_{LN}({\pi\over 2}),{\pi\over 2})\mid^2
\cot({{\omega T-N\Gamma}\over{2}}})}.
\eea

	As usual we assume that $\omega$ has a small positive
imaginary part so that $\omega$ in the equation above has to be
interpreted as $\omega +i\eta$. The poles of $\delta R_{LM}^0 (\omega)$
in the complex $\omega$-plane are determined by the vanishing of the
denominator in Eq. (\ref{dr0}). This condition can be written in more
compact form as
\beq
\label{main}
C_L -\chi_L  (\omega)=0
\eeq
with $C_L$ given by Eq.(\ref{cl}) and
\beq
\label{chi}
\chi_L (\omega) \equiv \omega~ R^2 {{(4\pi)^2}\over{(2L+1)}}
\int d\epsilon ({{\partial n_0}\over{\partial \epsilon}})
\int d\lambda \lambda p_{r}^2 
\sum_{N}\mid Y_{LN}({\pi\over 2},{\pi\over 2})\mid^2
\cot({{(\omega+i\eta) T-N\Gamma}\over{2}}).
\eeq
It is useful to recall that the functions
$Y_{LN}({\pi\over 2},{\pi\over 2})$ vanish unless $N$ has the same parity as
 $L$ so that the sum over $N$ involves only terms with either odd or even $N$.

	Eq.(\ref{main}) is the main result of Ref. \cite{abr} and its solution
determines the frequencies of collective nuclear excitations
within that approach.

\section {Comparison with other models}

	It is interesting to compare Eq.(\ref{main}) with the analogous condition
given by the LDM. We consider the version that allows also for compression
modes\cite{boh}.
Equation (6A-58) of Ref.\cite{boh} can be written as
\beq
\label{ldm}
C_L -\omega^2 D_L [{L\over R}{{j_L ({\omega\over u_c }R)}
\over{\partial\over \partial r}j_L ({\omega\over u_c }r)\mid_{r=R}}]=0
\eeq
with $u_c $ the velocity of sound.

	Equations (\ref{ldm}) and (\ref{main}) are coceptually similar
because they both express a condition for the eigenfrequencies of collective
nuclear modes. The main difference between them is that Eq. (\ref{ldm})
has been obtained in a macroscopic hydrodynamic approach,
while Eq.(\ref{main}) has been derived within a microscopic kinetic-equation
approach. It is thus reasonable to expect both similarities
and differences between the eigenmodes determined by the two equations.

	For small $\omega$ the term in square brackets in (\ref{ldm})
tends to 1 and Eq.(\ref{ldm}) becomes
\beq
C_L - \omega^2 D_L =0.
\eeq
In this approximation the surface vibrations described by $\delta R_{LM}(t)$
are pure harmonic oscillations and $\delta R_{LM}(\omega)$ has two simple poles
at $\omega_L =\pm\sqrt{{C_L \over D_L }}-i\eta $.

	In order to evaluate the function $\chi_L (\omega)$ we take
 $({{\partial n_0}\over{\partial \epsilon}})=
-{4\over(2\pi)^3}\delta (\epsilon_F -\epsilon)$, as appropriate for a gas of
zero-temperature nucleons ($\epsilon_F$ is the Fermi energy,
we use units such that $\hbar=1$),
	Even though Eq.(\ref{main}) is only a zero-order approximation,
it is interesting since it allows us to
obtain analytical expressions for the mass parameters $D_L$ 
and for the other dynamical coefficients that determine the
eigenfrequencies of collective modes. Unfortunately the functions
$\chi_L (\omega)$ are not simple, only $\chi_{L=0}$
has been evaluated explicitly for any
value of $\omega$. The result
derived in Ref. \cite{abr} is, in terms of the dimensionless parameter
$s\equiv {\omega \over (v_F /R)}$,
\beq
\label{mon}
\chi_{L=0}(s) =-2\epsilon_F \varrho_0 R^3 \{1+6
\sum_{n=1} ^{\infty}[{1\over 3}+{1\over{s_n ^2}}-{1\over{s_n ^4}}W(s_n)]\},
\eeq
with
\beq
\label{ws}
W(s)={1\over 2}s \ln \mid{{s+1}\over{s-1}}\mid +i{\pi\over 2}s~
\Theta(\mid s\mid -1),
\eeq
 $\varrho_0 ={{2p_F ^3}\over{3\pi^2}}$  the equilibrium density of
nuclear matter and $s_n =s/n\pi$.

	We remark that, contrary to the hydrodynamic relation (\ref{ldm})
that involves only real quantities, Eq. (\ref{main}) involves complex
quantities and thus we can expect complex eigenfrequencies as solutions,
in close analogy with the Landau damping phenomenon in homogeneous systems.

	In this paper we want to extend the study of the functions
$\chi_L (\omega)$, that was initiated in Ref. \cite{abr} for the monopole case,
to multipolarities $L>0$. For this purpose we write Eq.(\ref{chi}) as
\beq
\label{chi2}
\chi_L (\omega)=P_L ^2 (0)\chi_{L=0}(\omega) +{\cal N}_L \sum_{N>0}^L
\mid Y_{LN}({\pi\over 2},{\pi\over 2})\mid^2 I_N (\omega),
\eeq
with
\beq
\label{ins}
I_N (\omega)=-\omega\int_0 ^1 d\cos\alpha ~\cos\alpha \sin^2 \alpha
~[\cot z_N (\alpha) +\cot z_{-N}(\alpha)],
\eeq
\beq
z_N (\alpha)={{(\omega+i\eta)\sin\alpha}\over{v_F /R}}+N\alpha,
\eeq
and
\beq
{\cal N}_L ={2\over \pi^2}(p_F R)^4 {{4\pi}\over{2L+1}}.
\eeq
The quantities $P_L (x)$ are Legendre polynomials of order $L$. It is useful to
recall that
\bea
{{4\pi}\over{2L+1}}  \mid Y_{L0}({\pi\over 2},{\pi\over 2})\mid^2 =P_L ^2 (0)
&=& 0\qquad ~~~~~~~~{\rm odd}~L\nonumber \\
           &=& [{{L!}\over{2^L ({L\over 2}!)^2}}]^2 \quad {\rm even}~L
\eea

	To derive Eq.(\ref{chi2}) we have defined $\cos\alpha={{\lambda}
\over{p_F R}}$ in Eq.(\ref{chi}), the angle $\alpha$ has the simple
geometrical interpretation shown in Fig.1. It follows immediately that
$p_r =p_F \sin\alpha$, $T={{2R\sin\alpha}\over{v_F}}$, $\Gamma=2\alpha$.

	The integrals $I_N$ can
be complex because the integrands can have poles.
In order to evaluate these integrals it is convenient to use the following
trigonometric identity
\beq
\label{tid}
\cot z_1 +\cot z_2 ={{2\sin(z_1 + z_2)}\over{\cos(z_1 -z_2)-\cos(z_1 +z_2)}}
\eeq
and to change the integration variable to $x=\sin \alpha$, so that
\beq
I_N (\omega) =2\omega\int_0 ^1 dx x^3{{\sin(2xs)}
\over{\cos[2x(s+i\eta)]-\cos(2N\arcsin x)}}
\eeq
	The terms $\cos(2N\arcsin x)\equiv y_N$ are just polynomials in $x^2$:
$y_1 =1-2x^2 $, $y_2 =1-8x^2 +8x^4 $,
$y_3 =1-18x^2 + 48x^4 -32x^6 $, and so on.
The integrals $I_N$ can be easily
evaluated for small $s$, thus we are led to study the following 
low-frequency expansion for the functions $\chi_L (\omega)$:
\beq
\label{exp}
\chi_L (\omega) \approx A_L +i\omega \gamma_L +D_L \omega^2 .
\eeq
This expansion is valid for $s<1$, that is up to excitation energies of
about $46 A^{-1/3}$ MeV, however in the monopole case the upper limit
of validity is $\pi$ times larger, since in that case the relevant
expansion parameter is $s/\pi$, thus we can expect to describe also the
monopole compression mode within this approximation.

	The three coefficients $A_L$, $\gamma_L$ and $D_L$
given by the present approach can be compared with the same coefficients
obtained in the hydrodynamic approximation and to those derived in the more
closely related approach of Ref. \cite{koo}. As we shall see, some of these
coefficients vanish exactly, others (the mass parameters $D_L$) may even
diverge, in any case they
convey non trivial information about the low-frequency behaviour of the
functions $\chi_L (\omega)$ and $\delta R_{LM}(\omega)$.

	The coefficients $A_L$ in Eq.(\ref{exp}) vanish for odd multipoles,
while for even multipoles
\beq
A_L = -2\epsilon_F \varrho_0 R^3 P_L ^2 (0).
\eeq
The nonvanishing coefficients $A_L$ renormalize the restoring
force parameter $C_L$ in Eq.(\ref{main}). The largest value of $\mid A_L \mid$
occurs for $L=0$ when $A_0 =-2\epsilon_F \varrho_0 R^3 $. For increasing (even)
$L$ the coefficients $A_L$ tend to zero as $1/L$, while $C_L$ in Eq. (\ref{cl})
increases as $L^2$, so this renormalization becomes negligible in the limit of
large $L$. However for $L=0$ $A_L$ is much larger than $C_L$, in agreement
with the well known LDM result that the monopole excitation is not a surface,
but a compression mode (in this case the surface tension plays a negligible role).
For $L=2$ the value of $A_L$ is comparable to that of $C_L$
and this leads to a reduction of the role played by the surface
tension in this excitation mode too.

	The friction coefficients $\gamma_L$ are most easily evaluated directly
from Eq.(\ref{chi}) by using the pole expansion
\beq
\cot z=\sum_{n=-\infty}^{+\infty}{{1}\over{z-n\pi}}.
\eeq
It is clear also from Eq. (\ref{ws})  that for $L=0$ the friction
coefficient vanishes, while for $L>0$ we find
\beq
\label{gamma}
\gamma_L =\gamma_{wf}~2{(4\pi)^2\over{2L+1}}\sum_{N=1}^L {1\over N}
\mid Y_{LN}({\pi\over 2},{\pi\over 2})\mid^2
\sum_{n=1}^{+\infty} \cos\alpha_{nN}\sin^3 \alpha_{nN}
\Theta({\pi\over 2}-\alpha_{nN})
\eeq
with $\alpha_{nN}={n\over N}\pi$. The angles $\alpha_{nN}$ appearing in this
equation are related to the nucleon trajectory in the unperturbed mean
field in the way shown in Fig.1 . It is interesting to note that when the
coefficient $\gamma_L$ is
nonvanishing, it gets a contribution only from nucleons moving along
closed classical trajectories. Because of the vanishing of the coefficients
$Y_{LN}({\pi\over 2},{\pi\over 2})$ when the parity of $N$ differs from
that of $L$, for even multipoles the sum over $N$ in Eq.(\ref{gamma})
effectively starts from $N=2$ and involves only even values of $N$.

For ease of comparison with Ref. \cite{koo} we have expressed
our values of $\gamma_L$ in terms of the wall formula friction
coefficient \cite{blo}
$\gamma_{wf}\equiv {3\over 4}\varrho_0 p_F R^4$. We find that Eq.(\ref{gamma})
reproduces very well the values of $\gamma_L$ reported in Ref.\cite{koo}.
For all value reported there (up to $L=12$) we find perfect agreement
for odd multipoles, while for even multipoles our values of $\gamma_L$ are
systematically about $3-4\%$ smaller than the values of\cite{koo}. This
discrepancy tends to decrease with increasing $L$, and is only about
$1\%$ for $L=12$.
	
	From Eqs.(\ref{chi2}) and (\ref{ins}) we can obtain an explicit expression for the
coefficients $D_L$ in the expansion (\ref{exp}), it reads
\beq
\label{dls}
D_{L}={1\over 5}m\varrho_0 R^5 ~P_L ^2 (0)~ + 12m\varrho_0 R^5
{{4\pi}\over{2L+1}}\sum_{N>0}^L \mid Y_{LN}({\pi\over 2},{\pi\over 2})\mid^2
~I'_N\,
\eeq
with
\beq
I'_N \equiv \lim_{s\to 0}\int_0 ^1 dx{{x^2}
\over{{{1-\cos(2N\arcsin x)}\over{x^2}}-2s^2 -i\eta s}}.
\eeq
Note that possible terms that behave as $1/s$ resulting from the evaluation
of $I'_N$ should be omitted from $D_L$
since they have already been included in the coefficients $\gamma_L$ and that
the sign of the imaginary part of $I'_N$ depends on the sign of $s$.

The first term on the $r.h.s$ of Eq.(\ref{dls}) is present only for even
multipolarities and is the only term for $L=0$, the remaining sum over $N$
involves only even or odd values of $N$, depending on the parity of $L$.

	The first few integrals $I'_N$ are:
\beq
I'_1 = {1\over 6},
\eeq
\beq
\label{qua}
I'_2 =-{1\over 8}\lim_{s\to 0} \ln\mid s\mid+ 
{1\over 8}(2\ln{2}-1) + i{\pi\over 16}{\rm sign}~s
\eeq
\beq
I'_3 =-{1\over 64}{\sqrt{3}\over 3}(4\sqrt{3}+
\ln{{1+\sqrt{3}/2}\over{1-\sqrt{3}/2}}) +i
{\pi\over 16}{\sqrt{3}\over 2}\lim_{s\to 0}{1\over s}
\eeq
	For all even multipoles with $L>0$ the real parts of the integrals $I'_N$
are divergent because there is a pole exactly at the upper integration limit.
This divergence agrees with that found in Ref.\cite{koo}.
In the present formalism  it is possible to pinpoint exactly
the origin of this divergence: it is generated by nucleons moving on orbits
along the diameter. A peculiar feature of the present approach is that the
coefficients $D_L$ can be complex. In this aspect our results differ radically
from those of Refs. \cite{boh} and \cite{koo}. Our coefficients $D_L$
are real for $L=0,1$, but for $L=2$ the real part is divergent and the imaginary
part is finite. Note that this does not imply that the function $\chi_{L=2}$
has a pathological behaviuor for $\omega \to 0$, since the divergence is only
logarithmic.

 The integral $I'_3$ is also
complex, however its imaginary part behaves like $1/s$, the corresponding
term in the expansion (\ref{exp}) is linear in $\omega$ and it has already
been included in the coefficient $\gamma_L$. Thus for example for $L=3$ 
only the real part of $I'_N$ should be
included in $D_L$.

	In Table 1 we compare the values of the dynamic parameters given
by the present approach with those of Refs. \cite{boh,koo} for the first
few multipoles.
	
	As already found in Ref. \cite{abr}, the present approach gives the
following value for the energy of breathing modes in nuclei:
\beq
\label{mono}
\hbar \omega_{L=0} \approx\hbar\sqrt{{{C_0-A_0}\over{D_0}}} \approx
96~A^{-1/3}{\rm ~MeV}.
\eeq
This value can be compared with that of $65~A^{-1/3}{\rm ~MeV}$ reported
in Ref. \cite{boh} for classical hydrodynamics (both numbers have been
obtained by assuming the parametrization $R=1.2~A^{1/3}{\rm ~fm}$, for this
reason our value differs from the value $103~A^{-1/3}{\rm ~MeV}$ of Ref. \cite
{abr}). We recall that the result (\ref{mono}) is based on a zero-order
approximation in which mean-field fluctuations in the bulk have been
neglected, taking into account these effects leads to \cite{sme}
$\hbar\omega_0 \approx 82~A^{-1/3}{\rm ~MeV}$, in reasonable agreement
with experiment \cite{van}.

	The $L=1$ case is the only one in which all models of Table 1 give the
same results. This simply means that none of them has problems in describing
centre-of-mass motion.

	For $L=2$ the eigenfrequency condition (\ref{main}) cannot have
solutions corresponding to real, small, $\omega$ because the coefficient
$D_{L=2}$ is complex. Thus we have to look for possible solutions in the
complex $\omega$ or $s$ plane. For this puspose we write Eq.(\ref{chi}) as
\beq
\chi_{L=2}(s)=3A_0 s\{{1\over 4}\int_0 ^1 dx x^3 \cot(sx)
+{3\over 8}\int_0 ^1 dx x^3 [\cot(sx-2\alpha) +\cot(sx+2\alpha)]\},
\eeq
with $\alpha=\arcsin x$. For small $s$, $\cot(sx)\approx 1/sx -sx/3 +{\sl O}
(s^3)$, and the first integral is easily evaluated. Similarly, the second
integral, using the identity (\ref{tid}), becomes
\bea
& &\int_0 ^1 dx x^3 [\cot(sx-2\alpha) +\cot(sx+2\alpha)]\} \nonumber\\
&=&4s\int_0 ^1 dx{{x^2}\over{8(x^2 -1)+2s^2}}\quad +{\sl O}(s^3)\nonumber\\
&=&-{s\over 2}[{1\over 2}\ln(1-{16\over s^2}) -1] \quad +{\sl O}(s^3).
\eea
Thus for complex $s$ and $|s|<1$ we have
\beq
\chi_{L=2}(s)=A_0 \{{1\over 4} -{s^2 \over 20} -{9\over 16}s^2 [{1\over 2}
\ln({16\over s^2}-1)+i{\pi\over 2} {\rm sign}(\Re~s)-1]\}\quad +{\sl O}(s^4).
\eeq

	If we use this approximation for $\chi_2 (s)$ the eigenfrequency
condition (\ref{main}) has  no solution with $|s|<1$ in the complex $s$-plane,
thus the present model, at least in its simple-minded zero-order approximation,
does not seem to be adequate for describing low-frequency 
isoscalar quadrupole oscillations in nuclei. Calculations of the isoscalar quadrupole
response performed in Ref.\cite{kol} within the same model used here (for real $\omega$)
suggest the existence of a collective mode in a region where $|s|\approx 1$ (cf. Fig.2b of
\cite{kol}). Taking into account the attractive interaction inside nuclei (that is
the term $\delta U_{LM}$ in Eqs.(\ref{b6}) and (\ref{bf})) could bring this collective
mode into the region $|s|<1$, however this would require a more numerical approach
and would mean loosing the insight given by the analytic expressions of the zero-order
approximation.

	For $L=3$ the eigenfrequency condition (\ref{main}) in the limit of
small $\omega$ becomes
\beq
i\omega \gamma_3 +D_3 \omega^2=C_3.
\eeq
Using the values of parameters reported in Table 1, we find that low-energy
octupole oscillations are overdamped. This result might offer a qualitative
explanation for the background observed in inelastic proton scattering at small
angle, that has been previously studied by assuming a semi-infinite model
\cite{esb}, \cite{ran}.

\section{Summary and conclusions}

	Different solutions of the linearized Vlasov equation for finite
systems can be obtained, depending on the boundary conditions imposed
on the fluctuations of the single-particle phase-space density.

	We have discussed two different solutions
(with fixed and moving surface) that can be useful in studying collective
nuclear excitations. In particular we have analyzed the moving-surface
solution in the low-frequency limit for the first few isoscalar nuclear
modes (monopole, dipole, quadrupole and octupole) and have derived
interesting analytical expressions for the friction and mass parameters
entering the equation of motion for the oscillations of the nuclear surface.
According to our semiclassical result, in the low-frequency limit,
nuclear dissipation is due only to nucleons moving along closed classical
trajectories.
Numerically our friction parameters are in rather good agreement
with those derived in Ref. \cite{koo}. For the mass parameters our
result agree only partially with those of \cite{koo},
like in that case we find a divergence of the mass parameters for even
multipolarity, but our parameters can be complex, with a finite imaginary part.
For odd multipolarities (dipole and octupole) our mass parameters do agree
with those of Ref. \cite{koo}. Thus, even though the formalism 
is quite different, we conclude that the approach of Ref. \cite{abr},
is substantially equivalent to that of Ref. \cite{koo}.
On the contrary we find several differences with respect to the classical
hydrodynamic results of Ref. \cite{boh}. These differences, the most
imporant being probably the Landau damping of surface oscillations, can be
ascribed to the fact that the properties of a fluid of fermions are,
even in a semiclassical description,
different from those of a classical fluid. 

\section{acknowledgements}

	V. A. acknowleges financial support and kind hospitality from
INFN, Italy.

%%%%%%%  REFERENCES  %%%%%%%%%%%%%%%%%%%%%%%%%%%%%%%%%%%%%%%%%%%%%%%%%%%%%%%
%

%%%%%%%%%%%%%%%%%%%%%%%%%%%%%%%%%%%%%%%%%%%%%%%%%%%%%%%%%%%%%%%%%%%%%%%%%%%%%
\begin{figure}[htb]
\centerline{\epsfig{figure=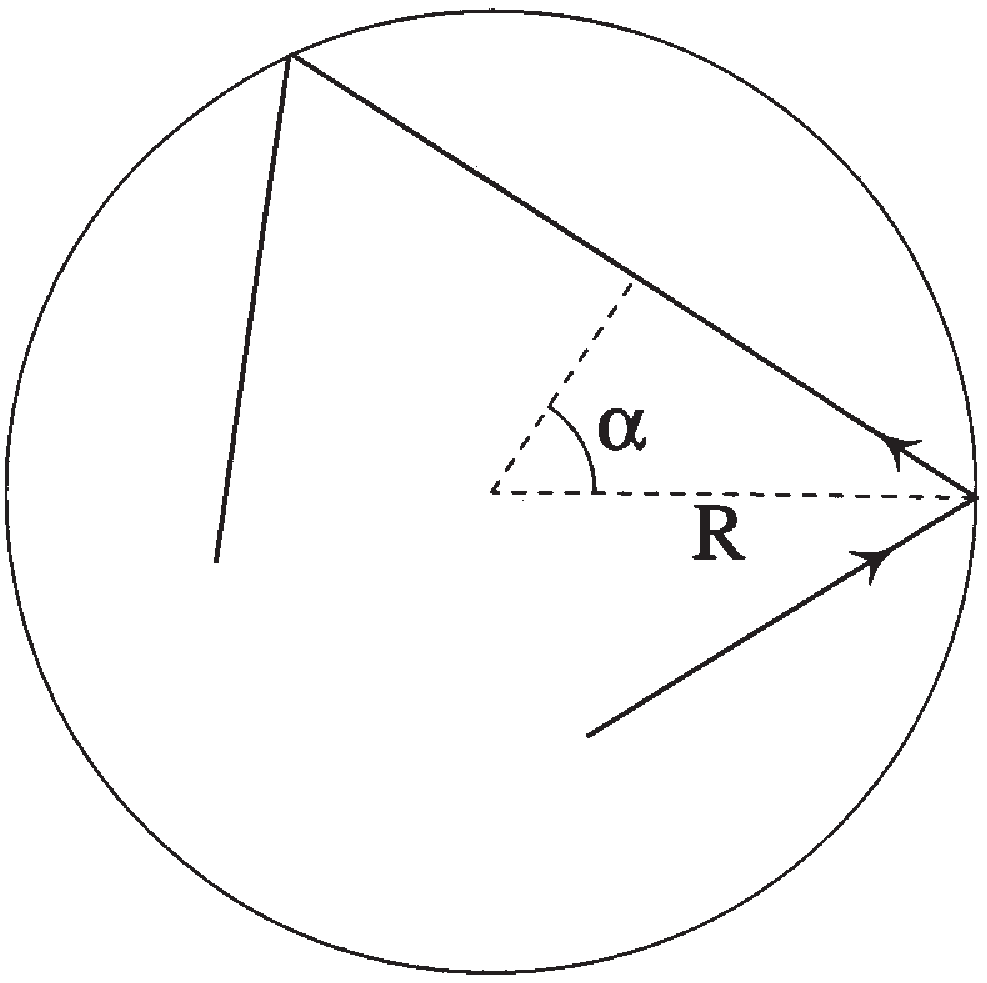}}
\caption[ ]{Relation of angle $\alpha$ to nucleon trajectory in equilibrium
mean field}
\end{figure}

\begin{table}
\caption{Dynamic parameters for lowest multipolarities.\hfill\break
Columns a: present results, columns b: Ref.[7], columns c:
model of Ref.[6]}
\label{t:1}
\begin{tabular}{ccccccccc}
$L$ & \multicolumn{2}{c}{$A_L /A_0$} 
&\multicolumn{3}{c}{$\gamma_L/\gamma_{wf}$}
&\multicolumn{3}{c}{$D_L/m\varrho_0 R^5$}             \\
    & a    & c     & a	   & b	   & c	 & a	 & b	    &c        \\
\tableline
0   & 1    & 0     & 0       & 0    & 0	 &0.2	 & $\infty$ &       --  \\
1   & 0    & 0     & 0       & 0    & 0   &1     &1         &1           \\
2   & 0.25 & 0     & 0       & 0    & 0   &$+\infty +i{9\over 32}\pi$ &$\infty$
 &0.5 \\
3   & 0    & 0     & 0.85    & 0.85 & 0   &0.052  &0.050      &0.333 \\ 
\end{tabular}
\end{table}
\end{document}